\documentstyle[11pt,epsfig]{article}
\title{{\bf Observational Evidence for Galaxy Evolution in the Local Group}}
\author{Eline Tolstoy$^1$\\
\vspace{0.1cm}\\
\normalsize $^1$UK Gemini Support Group, University of Oxford, Keble Rd, Oxford OX1 3RH, UK\\
}
\date{}
\begin{document}
\maketitle
\def\bull{\vrule height .9ex width .8ex depth -.1ex}
\makeatletter
\def\ps@plain{\let\@mkboth\gobbletwo
\def\@oddhead{}\def\@oddfoot{\hfil\tiny
``Dwarf Galaxies and their Environment'';
International Conference in Bad Honnef, Germany, 23-27 January 2001}%
\def\@evenhead{}\let\@evenfoot\@oddfoot}
\makeatother

\begin{abstract}\noindent
This review aims to give a summary of our understanding of galaxy
evolution as infered from studies of nearby galaxies; how observations
made with the {\it Hubble Space Telescope} have contributed
significantly to our detailed understanding of the older stellar
populations in Local Group dwarf galaxies.  Recent results from VLT
are also promising interesting future prospects for the study of
resolved stellar populations in nearby dwarf galaxies.
\end{abstract}

\section{Introduction}

How do galaxies form and then evolve over time? This is one of the
most fundamental questions in astronomy, and the answer has far
reaching implications for the accurate interpretation of any
observations of galaxies throughout the Universe.  Galaxies are the
end products of all the star formation in their entire lifetimes, and
the ratios of chemical elements and remaining stellar population
provide evidence for this past star formation.  Only if we understand
how galaxies change with time and especially how they may look when
they are young can we use them to accurately understand what we see in
galaxy surveys at high-redshift, because otherwise we don't know which
type of galaxies we may be viewing.  There are currently numerous
techniques available to uncover information buried in the properties
of individual stars.  It will make a significant difference in the
interpretation of galaxy surveys if they preferentially detect
populations of star-bursting dwarf galaxies - which do not trace the
mass distribution in the universe - or if the galaxies we detect are
massive spirals or ellipticals which do.  We thus have to understand
how different types of galaxies evolve so we are able to distinguish
their progenitors in redshift surveys.

It must be fair to assume that all the galaxies we see today in and
around our Local Group are broadly representative; our region of space
is neither over-dense (a cluster environoment) nor under-dense (in a
void). All nearby galaxies have doubtless been forming and evolving
for a significant fraction of the age of the Universe.  If this were
not true it would mean that our local region of space is in some way
peculiar, and there is no evidence for this.  Thus, as new techniques
and instruments enable us to determine more and more accurate star
formation histories for nearby galaxies over 90\% of the lifetime of
the Universe we can hope to obtain a representative picture of galaxy
evolution from our local neighbourhood, and with it the ability to
predict what galaxies look like at all redshifts.  Detailed studies of
nearby galaxies will thus provide an independent method to compare
with redshift survey predictions.

It is apparent that some galaxies have a more or less constant global
star formation rate through time ({\it e.g.}  spirals) and some appear
to be subject to sudden, intense {\it bursts} of star formation ({\it
e.g.} irregulars) and then some stopped forming stars entirely at some
point in the past ({\it e.g.}  ellipticals).  Filling in the crucial
details of this basic scenario requires the detailed analysis of the
fossil record of ancient star formation.

\begin{figure}[h]
\begin{center}
\epsfig{file=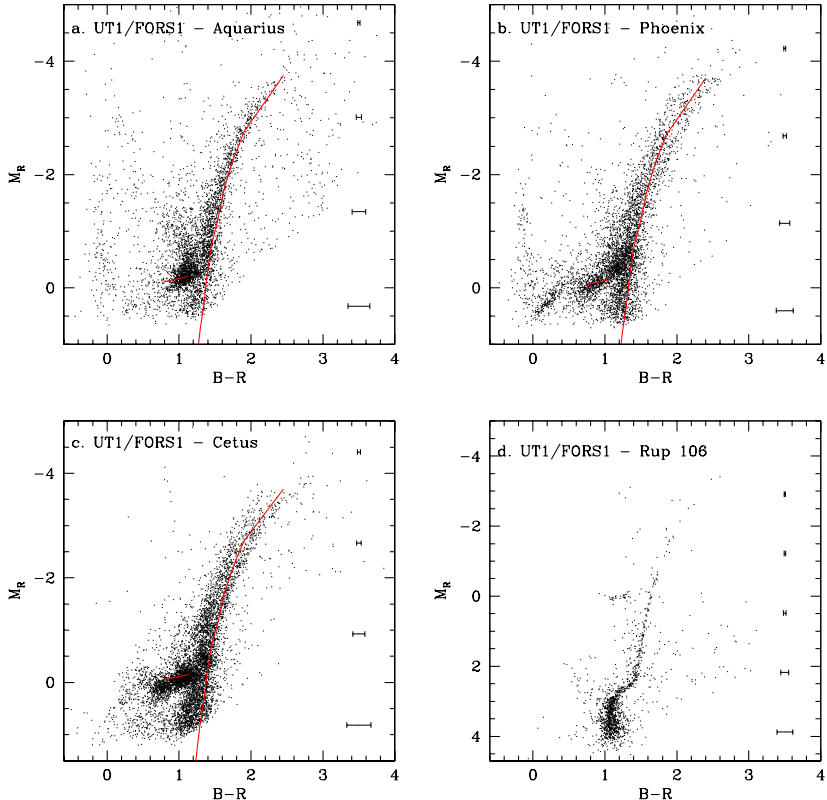,width=12cm}
\caption[]{
Here are plotted the Colour-Magnitude Diagrams for Local Group Dwarf
galaxies (a) Aquarius, (b) Phoenix and (c) Cetus, and as a comparison,
(d) the globular cluster Ruprecht 106, which resulted from UT1/FORS1
imaging in August 1999, in exceptional seeing conditions (Tolstoy {\it
et al.} 2000).  Representative error bars are also plotted for each
data set.  These data have not been corrected for any reddening
effects. From the Ruprecht~106 data a fiducial mean was found for the
RGB and HB. This is then over-plotted on each of the dwarf galaxy CMDs.
}
\end{center}
\end{figure}

\section{Imaging: Star-Formation Rate Evolution}

The most detailed information on how a galaxy has evolved in time
comes from measuring the star-formation rate as a function of time, or
the star formation history (SFH), and the most direct and unambiguous
method of doing this comes from interpreting CMDs of a significant
fraction of the individual stars in a galaxy. This is a plot of the
temperature versus the luminosity of all the stars bright enough to be
detected in a galaxy in the observed quantities of colour and
magnitude (see Figure~1).  Because of our detailed understanding of stellar
properties these measurements can be converted into physical
parameters such as age (or SFH), chemical composition and enrichment
history, initial mass function, environment, and dynamical history of
a system.  Some of these parameters are strongly correlated, such as
chemical composition and age, since successive generations of stars
may be progressively enriched in the heavier elements.  Thus, detailed
numerical simulations of CMD morphology are necessary to disentangle
the complex effects of different stellar populations overlying each
other and make an effective quantitative analysis of possible SFHs. To
this end There have been numerous recent developments in crowded field
photometry techniques and advanced methods of interpreting highly
populous and detailed CMDs from nearby galaxies (e.g., Tosi {\it et al.}
1991; Tolstoy \& Saha 1996; Aparicio {\it et al.} 1996; Dolphin 1997;
Dohm-Palmer {\it et al.} 1997; 
Hernandez {\it et al.} 1999; Harris \& Zaritsky 2001).

\begin{figure}[t]
\begin{center}
\epsfig{file=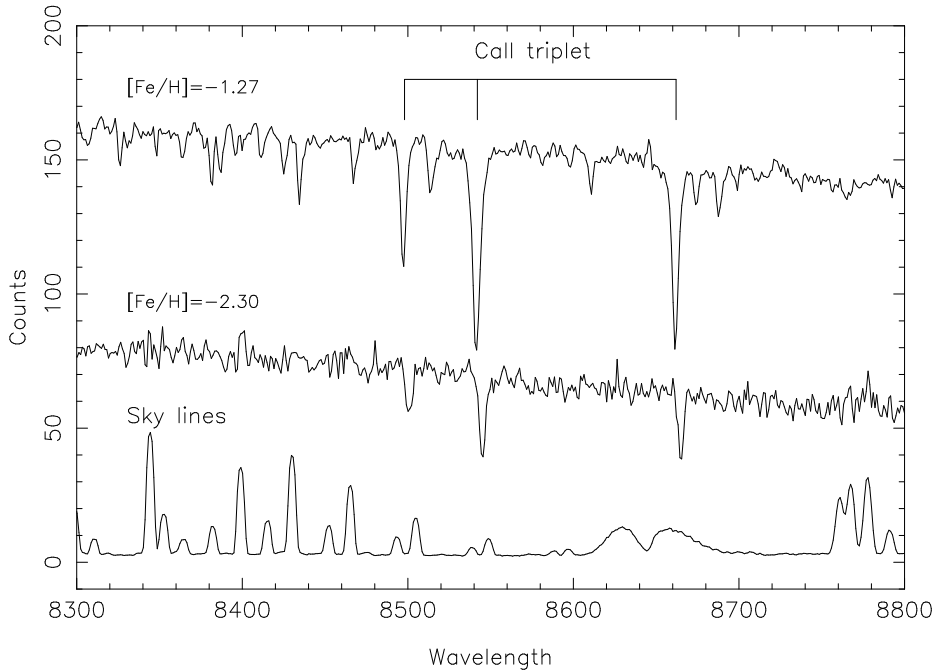,width=10cm,angle=-90}
\caption[]{
Here we show two examples of UT1/FORS1 low-resolution spectra of the
Ca~II triplet region of two red giant stars in the Sculptor and Fornax
dwarf spheroidal galaxies observed in August 1999 (Tolstoy {\it et
al.}  2001). They both lie at the opposite extremes of Ca~II triplet
line widths measured in our sample.  For display purposes the spectra
have been normalized to their continuum level and then arbitrarily
shifted.  The upper spectrum is of star c2-838 in Fornax with a
calcium triplet metallicity of [Fe/H]$= -1.27$, and the lower spectrum
is of star o1-1 in Sculptor, with [Fe/H]$= -2.30$.  They both have
good S/N, with $\sim$30 in the upper spectrum and $\sim$20 in the
lower.  Also shown here is the sky spectrum. This shows that, although
this region of the spectrum is relatively free of bright sky lines,
the weaker Ca~II triplet line at 8498\AA \, is more likely to be
affected by sky lines than the other two.
}
\end{center}
\end{figure}

The exquisite stable high spatial resolution combined with photometric
accuracy of images from the {\it Hubble Space Telescope} (HST) have
allowed us to probe further back into the history of star formation of
a large variety of different galaxy types with widely differing star
formation properties, and extend our studies out to the edges of the
Local Group and beyond with greater accuracy than ever before.  We
have learnt several important things about dwarf galaxy evolution from
these studies. Firstly we have found that no two galaxies have
identical star formation histories; some galaxies may superficially
look the same today, but they have invariably followed a different
paths to this point. Now that we have managed to probe deep into the
star formation history of dwarf irregular galaxies in the Local Group
it is obvious that there are a number of similarities in their global
properties with those of dwarf elliptical/spheroidal type galaxies,
which were previously thought to be quite distinct. However, the
elliptical/speheroidals tend to have one or more discrete episodes of
star formation through-out their history and dwarf irregulars are
characterised by quasi-continuous star-formation. The previous strong
dichotomy between these two classes has been weakened by these new
results and may stem from the differences in the environment in which
these similar mass galaxies were born into or have inhabited for most
of their lives.

\begin{figure}[h]
\begin{center}
\epsfig{file=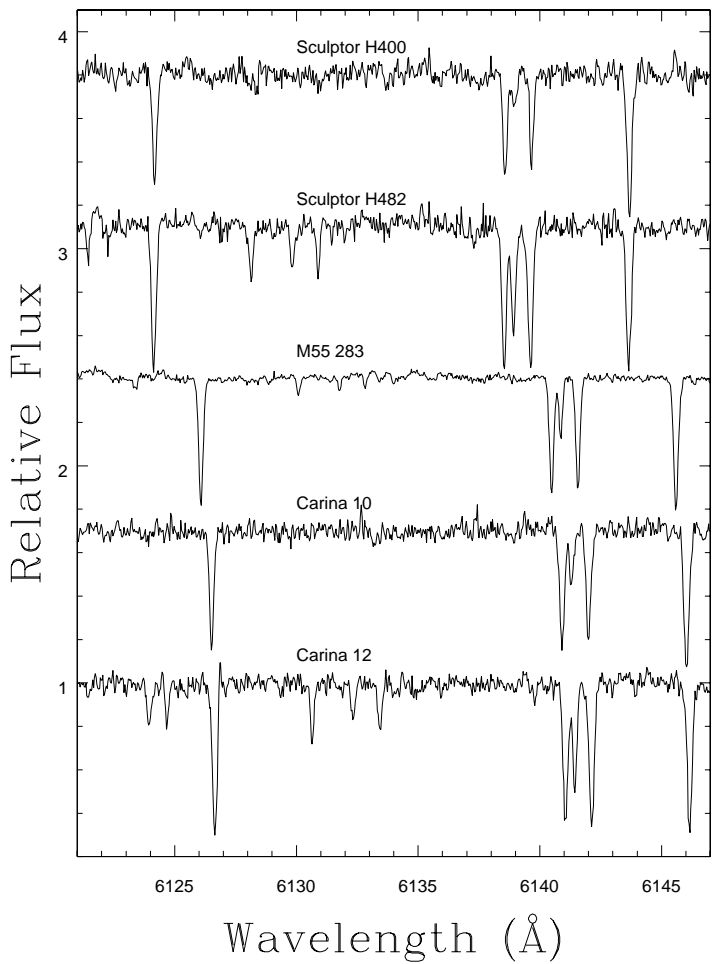,width=10cm}
\caption[]{
Here we show a small region of four UT2/UVES high-resolution spectra
observed in August 2000 and January 2001 of stars in Sculptor and
Carina dwarf spheroidals, and one of a comparison star in metal-poor
globular cluster M~55 (Tolstoy, Venn, Shetrone, Hill, Primas, Kaufer
\& Szeifert in prep).  
Here we can see lines from Ca~I
($\lambda\lambda$6122.2) Fe~I ($\lambda\lambda$6136.6, 6137.7) and
Ba~II ($\lambda\lambda$6141.7).  Clearly both Carina and Sculptor
exhibit a range of abundance variation within their stellar population,
and there are also interesting difference between the two galaxies.
}
\end{center}
\end{figure}

\section{Spectroscopy: Metallicity Evolution}

If we want to understand the detailed chemical enrichment of galaxies
by their constituent stars, we need to accurately measure the relative
abundances of a range of different elements in high resolution
spectra.  To date most metallicity determinations for stars in nearby
galaxies (and there aren't many) have been derived either from single
chemical element studies from low resolution spectra (see Figure~2) or
from broad-band photometry, and these are at best simple estimates.
Thus in spite of considerable efforts with 4m class telescopes, many
of the most basic questions remain unanswered for most of the Local
Group galaxies: What is the total range in [Fe/H] for stars within a
galaxy?  Does the [Fe/H] match that of the Milky Way halo populations
(Globular clusters or field stars)? Do the elemental ratios
(particularly [$\alpha$/Fe] and [r- and s- process/Fe] ratios) in
nearby systems track those seen in the metal poor field or in the
globular clusters?

Results from high resolution spectra taken with HIRES at the Keck
Observatory, have started to answer these detailed questions about the
enrichment {\it history} of a variety of different elements within
galaxies other than our own (e.g. Shetrone {\it et al.} 1998; Shetrone
{\it et al.} 2001; Venn {\it et al.} 2001).  The new ESO {\it Very
Large Telescope} (VLT) observatory on Paranal in Chile consists of
four 8m diameter mirror telescopes and a slew of modern
instrumentation.  The large collecting area and the high resolution
spectrograph (UVES) allows us, amongst other things, to
spectroscopically determine the abundances of a large variety of
elements for individual stars (e.g., Hill {\it et al.} 2000; Primas
{\it et al.} 2000). Recent results of UVES spectroscopy of stars of
known age in a dwarf galaxy CMD are shown in Figure~3.  
By looking at galaxies outside our own we gain the advantages
of perspective and thus we have a better chance to build up a 
complete picture of metallicity variations over time.
The most interesting stars, to get a large range in age, are
faint enough in external galaxies to require an 8m telescope.
Including chemical evolution history in CMD analysis will result in a
significant improvement in our understanding of how galaxies evolve.

For example, rather than just measuring a single present day
end-product abundance of an element, we can select individual stars of
different ages from imaging data and {\it measure} how the enrichment
of many different elements has varied with time.  This means we can
measure how the chemical composition of the interstellar medium, the
basic building material for future stellar populations, is altered by
successive generations of stars.  Deep, precision, multi-colour
photometry in combination with spectroscopic metallicity measurements
of the individual stars in external galaxies can uniquely determine
the star formation histories of nearby galaxies going back many
giga-years.  These studies make clear the potential of deeper data,
from VLT and new HST instruments, for a range of galaxy types.

\section{Prospects for the future}

We are currently (apparently) in the embarrassing situation of, at least 
claiming to, understand the properties of distant high-redshift galaxies 
better than those in the nearby Universe.  It is only with the arrival of 
large telescopes with excellent image quality and high through-put 
spectrographs that we can start to make really detailed comparisons between 
the properties of distant and nearby galaxies.  This is because the 
individual low mass (old) stars in the nearby universe which formed when 
the Universe was young (at high redshift) are faint.  Thus to see them in 
galaxies external to our own requires at least an 8m class telescope. 

There are a number of ways in which the current interpretation of CMDs
can be dramatically improved.  One is more and deeper CMDs of a larger
sample of nearby galaxies. It is still most efficient to use HST to
obtain deep CMDs to study the old main sequence turnoffs, but large
scale surveys, down the magnitude of the Horizontal Branch can better
be carried out with ground-based telescopes such as the VLT (see
Figure~1).  When galaxies are satellites of our own Milky Way, they
are typically close enough that excellent results can best be achieved
using wide field imagers on 2m class telescopes (e.g. MPA/ESO/2.2m WFI
on La Silla or the INT/WFC on La~Palma), of which there have been some
nice results shown at this conference.  I believe that the most
important contributions will come from continuing the effort to
improve our ability to interpret the details in CMDs. There will be
significant progress when we have measured abundances of a large
variety of elements for stars of known age for individual stars in a
CMD.

The observed redshift distribution of faint galaxies detected in deep
UV/optical imaging surveys has been assumed to trace the star
formation history of the Universe 
(e.g., Madau {\it et al.} 1996; Lilly {\it et al.} 1996).  
The majority of these galaxies are
at intermediate redshift ($z<1$), late type, intrinsically small, and
undergoing a strong ``burst'' of star formation.  This means that 
the nearby Universe (the Local Group) must contain clearly
identifiable remnants from this relatively recent epoch $\sim5-8$~Gyr
before present (i.e.  corresponding to $z\sim$~0.5$-$1) when the peak
in the universal star formation rate is predicted to occur.  Initial
comparisons suggest that studies of nearby galaxies {\it do not} yield
the same SFH as is inferred from optical redshift surveys ({\it e.g.},
Tolstoy 1998).  However, the best present day candidates for this
intermediate redshift galaxy population are the extremely numerous but
presently faint (dwarf) irregular galaxies which have not yet been
studied in sufficient detail.  It is also true to say that although we
know that nearby spirals and especially ellipticals have very large
old stellar populations it is difficult to be very precise beyond
about $8-10$~Gyr ago (corresponding to a redshift range z$> 1-2$) with
current data.  This is all going to change dramatically with large
telescopes and more sensitive instruments.  Sub-mm/radio wavelength
surveys are also pointing to SFHs very different from those implied by
optical data (e.g., Blain {\it et al.} 1999), and 
these surveys suggest a much larger type of galaxy
({\it e.g.}, ellipticals) which are typically at higher redshift
($z>2$). 

A complete survey of the resolved stellar populations in
the local Universe will accurately trace star formation variations
within both large and small galaxies, and determine if and when bursts
of star formation occur and how long they last.  It might well be
that optical redshift surveys are strongly biased towards low mass
dwarf irregular galaxies undergoing short bursts of star formation,
and that they are thus not accurate indicators of the dominant mode
(by mass) of star formation in the Universe which occurs in much
larger galaxies.

The more detailed is our understanding of star formation processes and
their effect on galaxy evolution in the nearby Universe the better we
will understand the results from studies of the integrated light of
galaxies in the high-redshift Universe.

{\small
\begin{description}{} \itemsep=0pt \parsep=0pt \parskip=0pt \labelsep=0pt
\item {\bf References}
\item
Aparicio, A., Gallart, C., Chiosi, C. \& Bertelli, G. 1996, ApJL, 469, 97
\item
Blain, A. W., Smail, I., Ivison, R. J. \& Kneib, J.-P. 1999, MNRAS, 302, 632
\item
Dohm-Palmer, R. C., Skillman, E. D., Saha, A., Tolstoy, E., Gallagher, J. S., 
Hoessel, J. G., Mateo, M., \& Chiosi, C. 1997, AJ, 114, 2527
\item
Dolphin, A. 1997, New Astron. 2, 397
\item
Harris, J. \& Zaritsky, D. 2001, ApJS, in press
\item
Hernandez, X., Valls-Gabaud, D. \&  Gilmore, G. 1999, MNRAS, 304, 705
\item
Hill, V., Fran\c{c}ois, P., Spite, M., Primas, F. \& 
Spite, F. 2000, A\&A, 364, 19L
\item
Lilly, S. J., Le F\`{e}vre, O., Hammer, F., \& Crampton, D. 1996, ApJL, 460, 1
\item
Madau, P., Ferguson, H. C., Dickinson, M., Giavalisco, M., Steidel, C. C., 
\& Fruchter, A. 1996, MNRAS, 283, 1388 
\item
Primas, F., Molaro, P., Bonifacio, P. \& Hill, V. 2000, A\&A, 362, 666
\item
Shetrone, M. D., Bolte, M. \& Stetson, P. B. 1998, AJ, 115, 1888
\item
Shetrone, M. D., C\^{o}t\'{e}, P. \& Sargent, W. L. W. 2001, ApJ, 548, 592
\item
Tolstoy, E. 1998, in ``Dwarf Galaxies \& Cosmology'', 
eds T.X.~Thuan {\it et al.}, p. 171
\item
Tolstoy, E. \& Saha, A. 1996, ApJ,
\item
Tolstoy, E., Gallagher, J.S., Greggio, L., Tosi, M.,
De Marchi, G., Romaniello M., Minniti D. \& Zijlstra A. 2000, 
``ESO Messenger'', 99, 16 
\item
Tolstoy, E., Irwin, M. J., Cole A. A., Pasquini L., Gilmozzi R. \&
Gallagher J.S. 2001, submitted to MNRAS.
\item
Tosi, M. M., Greggio, L., Marconi, G. \& Focardi, P. 1991, AJ, 102, 951
\item
Venn, K. A., Lennon, D. J., Kaufer, A., McCarthy, J. K., Przybilla, N., 
Kudritzki, R. P., Lemke, M., Skillman, E. D. \& 
Smartt, S. J. 2001, ApJ, 547, 765
\end{description}
}
\end{document}